\documentclass[aps,preprint]{revtex4}
\usepackage[dvips]{graphicx}
\begin{document}
\draft
\title{\bf Experimental investigation of the elastic enhancement factor \\ in a transient region between regular and chaotic dynamics }

\author{Micha{\l} {\L}awniczak, Ma{\l}gorzata Bia{\l}ous, Vitalii Yunko, Szymon Bauch,  and Leszek Sirko}
\address{Institute of Physics, Polish Academy of Sciences,
Aleja \ Lotnik\'{o}w 32/46, 02-668 Warszawa, Poland \\}

\date{February 24, 2015}

\bigskip

\begin{abstract}

We present the results of an experimental  study of
the elastic enhancement factor $W$ for
a microwave rectangular cavity simulating
a two-dimensional  quantum billiard in a transient region between regular and chaotic dynamics. The cavity was coupled to a vector network analyzer via two microwave antennas. The departure of the system from the integrable one due to presence of antennas acting as scatterers is characterised by the parameter of chaoticity $\kappa = 2.8$.  The experimental results for the rectangular cavity are compared with the ones obtained for a microwave rough cavity simulating a chaotic quantum billiard. The experimental results were
obtained for the frequency range $\nu =16-18.5$ GHz and moderate
absorption strength $\gamma = 5.2-7.4$. We show that the elastic enhancement factor for the rectangular cavity lies below the theoretical value $W=3$ predicted for integrable systems  and it is significantly higher than the one obtained for the rough cavity.
 The results obtained for the microwave rough cavity are smaller than the ones obtained within the framework of Random Matrix Theory and lie between them and the ones predicted within a recently introduced model of the two-channel coupling (V. Sokolov and O. Zhirov, arXiv:1411.6211v2[nucl-th], 12 Dec 2014).
\end{abstract}

\pacs{05.45.Mt,03.65.Nk}
\bigskip
\maketitle

\smallskip

\section{Introduction}

The elastic enhancement factor was introduced more than 50 years ago by Moldauer \cite{Moldauer1961} and since then it has been
 frequently considered in nuclear physics \cite{Kretschmer1978,Verbaarschot1986,Kharkov2013} and in other fields \cite{Fyodorov2005,Savin2006}.
 The elastic enhancement factor $W_{\beta}$ is the ratio of variances of diagonal elements of the two-port
scattering matrix $\hat{S}$ to off-diagonal elements of this matrix \cite{Fyodorov2005,Savin2006,Kharkov2013}.
From the experimental point of view
the elastic enhancement factor $W_{\beta}$, where $\beta =1$ or $2$ is the symmetry index
for systems with preserved and broken time reversal symmetry, respectively,
is especially interesting because it can be used to study realistic open systems also in the presence of absorption.
The properties of the elastic enhancement factor $W_{\beta}$ have been studied in several precisely controllable systems
such as microwave cavities \cite{Fiachetti2003,Zheng2006,Dietz2010,Anlage2013} and networks
\cite{Lawniczak2010,Lawniczak2011,Lawniczak2012}.
The conjecture on the universality of the ratio of variances of the scattering elements
in electromagnetic fields in the mode-stirred reverberating chambers (time reversal invariant system)
 was put forward by Fiachetti and Michelson \cite{Fiachetti2003}.
 The universality of the elastic enhancement factor $W_{\beta=1}$ has been also tested
 in the wave scattering experiments with microwave cavities simulating
 chaotic quantum billiards \cite{Zheng2006,Dietz2010} in the presence of
absorption.
Dietz et al. \cite{Dietz2010} have studied
 the universality of the elastic enhancement factor $W_{\beta}$ with microwave cavities in the case of preserved
 and partially broken time reversal symmetries. Quite recently an extensive study
 of the elastic enhancement factor $W_{\beta=1}$ has been published by Yeh et al. \cite{Anlage2013}.
 In that paper the authors
 were also able to study the elastic enhancement factor for microwave cavities with time reversal symmetry in a low absorption regime.
The reciprocal quantity
$\Xi= 1/W_{\beta=1}$ was considered theoretically
and measured as a function of frequency for a chaotic microwave
cavity with time reversal symmetry  \cite{Zheng2006,Anlage2013}.

 The elastic enhancement factor $W_{\beta}$ has been also studied for microwave
  irregular networks \cite{Hul2004,Hul2012}  simulating quantum
graphs with preserved  and broken time reversal symmetry in
the presence of moderate and large absorption strength defined as follows: $\gamma =2\pi \Gamma /\Delta$,
 where $\Gamma $ is the average resonance width and $\Delta$ is the mean level
spacing \cite{Fyodorov2005,Savin2006}, $5\leq \gamma \leq 54.4$ \cite{Lawniczak2010,Lawniczak2011,Lawniczak2012}.
  Microscopically, the absorption strength $\gamma =\sum_c T_c$ can be modeled by means of a huge number of open, coupled to continuum channels "c", where $T_c = 1-|<S_{cc}>|^2$ and $<S_{cc}>$ stands for the average $S$-matrix \cite{Savin2006}. 
  The recent paper of Kharkov and Sokolov \cite{Kharkov2013} has shown that the elastic enhancement factor of open systems with a transient from the regular to chaotic internal dynamics depends on both the parameter of chaoticity $\kappa $ and
 the openness $\eta $. The openness $\eta $ is formally described by the same formula as the absorption strength  $\gamma $ \cite{Kharkov2013}.

It is important to point out that the elastic enhancement factor for the systems with absorption, in a transient region between regular and chaotic dynamics,  has not been studied experimentally yet.
In this paper we present the results of the experimental  study of
the elastic enhancement
factor $W_{\beta }$ \cite{Fyodorov2005,Savin2006}
for  microwave rectangular and rough cavities, coupled to the vector network analyzer through antennas,  simulating respectively,
partially chaotic and chaotic two-dimensional (2D) quantum billiards with preserved
time reversal symmetry ($\beta =1$) in the presence of moderate  absorption.

The elastic
 enhancement factor  $W_{\beta}$ is defined by the relationship \cite{Fyodorov2005,Savin2006}
\begin{equation}
\label{Eq.1}
W_{\beta }=\frac{\sqrt{\mbox{var}(S_{aa})\mbox{var}(S_{bb})}}{\mbox{var}(S_{ab})},
\end{equation}
where $\mbox{var}(S_{ab}) \equiv \langle |S_{ab}|^2\rangle
-|\langle S_{ab} \rangle |^2$ is the variance of the
scattering matrix element $S_{ab}$ of the
 two-port scattering matrix
\begin{equation}
\label{Eq.2}
\hat{S}=\left[
\begin{array}{c c}
S_{aa} & S_{ab}\\
S_{ba} & S_{bb}
\end{array}
\right]\mbox{.}
\end{equation}

For small and intermediate values of the parameter $\gamma$ the elastic
enhancement factor $W_{\beta}$ might depend both on the parameter $\gamma$ and
on the coupling to the system  \cite{Zheng2006}.
However, for large absorption strength $\gamma \gg 1$ the elastic
 enhancement factor  $W_{\beta}$ can be approximated by the formula:
$W_{\beta }=2/\beta $ \cite{Fyodorov2005,Savin2006,Zheng2006}.
Fiachetti \cite{Fiachetti2008} showed that in the case of the stochastic environment
which can be characterized by a statistically isotropic scattering matrix
the elastic enhancement factor
should have the universal value $W_{\beta=1}=2$. Recently, the two-channel problem (e.g., an experimental system with two ports "a" and "b") with internal absorption and time reversal symmetry has been numerically considered by Sokolov and Zhirov \cite{Sokolov2014}. It has been shown that for the equivalent channels "a" and "b" with the transmission coefficients $T_a=T_b=T$, $0\leq T \leq 1$, the elastic enhancement factor $W_{\beta=1}$ depends both on the transmission coefficient $T$ and internal absorption and  can take, respectively, the values between 3 and 2.
Thereafter, we will use the abbreviation $W \equiv W_{\beta=1}$.

\section{Microwave cavities simulating quantum billiards}

In the experiment we used a microwave  rectangular cavity to simulate a two dimensional (2D)
billiard in a transient region between regular and chaotic dynamics.  A quantum chaotic billiard was simulating by a rough microwave cavity.
If the excitation frequency $\nu $ is below $\nu_{max}=c/2d$,
where $c$ is the speed of light in the vacuum and $d$ is the height of the cavity,
only the transverse magnetic $TM_0$ mode can be excited inside the cavity. Then, the analogy between microwave flat cavities and quantum billiards
is based upon the equivalency of  the Helmholtz equation
describing the microwave cavities and the Schr\"odinger equation
describing the quantum systems \cite{Stockmann1990,Blumel}.

Absorption of the cavities can
either be changed by changing the frequency range of the measurements,
 or more effectively, by the application of microwave
absorbers. In this paper we are only interested in moderate absorption, for which $W>2$ \cite{Fyodorov2005,Savin2006,Zheng2006}
 and which
can be controlled by the choice of the microwave frequency range.

\section{Experimental setup}

 Figure~1(a)
 shows the scheme of the rectangular microwave cavity which was  used for measuring of the two-port scattering matrix $\hat{S}$.
The scattering matrix $\hat{S}$ of the cavity was measured in
the frequency window: 16--18.5 GHz.  The vector network analyzer
Agilent E8364B was connected  through the  HP 85133-616 and HP 85133-617
flexible microwave cables to the two microwave antennas which were introduced inside the cavity (holes $A_1$, $A_2$, $A_3$, $A_4$, $A_5$
in Figure~1(a)). The antennas wires (diameter 0.9 mm) were protruded 3 mm into the cavity. The measurements were completed for 10 different positions of the antennas.
 The width of the rectangular cavity was
$L_2=20$ cm. Different realizations of the cavity were created by the change of its length from $L_1 =
41.5$ to 36.5 cm in 25 steps of 0.2 cm length.

Figure~1(b) shows the scheme of the rough microwave cavity \cite{Hul2005}.
The cavity is composed of the two side wall segments. The segment (1) is
described by the function
$r(\theta)=r_{0}+\sum_{i=2}^{M}{a_{i}\sin(i\theta+\phi_{i})}$,
where  the mean radius $r_0$=20.0 cm, $M=20$, and $0\leq\theta<{\pi}$. The amplitudes $a_{i}$ and the phases
$\phi_{i}$ are uniformly distributed on [0.084,0.091] cm and
[0,2$\pi$], respectively.

Both, the rectangular and rough cavities had the same height $d=8$ mm, so that $\nu_{max}=18.7$ GHz.

 Also in the case of the rough cavity the two-port scattering
matrix $\hat{S}$ was measured in the frequency range 16--18.5 GHz. The 3 mm long antennas were introduced inside the cavity trough the holes $A_1$ and $A_2$.
 In order to create different realizations of the rough cavity a metallic perturber with the area $A_p \simeq 9$ cm$^2$ and the perimeter $P_p \simeq 26$ cm (see panel (b)) was moved inside the cavity along the sidewalls using an external magnet. The linear size of the perturber $L_p \simeq 5$ cm was more than 2.5 times bigger than the microwave wavelength at 16 GHz.

\section{The nearest neighbor
spacing distributions}

  Properties of
microwave cavities were investigated using the nearest neighbor
spacing distribution $P(s)$. The distribution $P(s)$ for the microwave
rectangular cavity obtained for the frequency range $\nu=16-17$ GHz is shown in Figure~2(a) (bars).  The
distribution $P(s)$ was averaged over 30 microwave cavity
configurations. In this way 2760 eigenfrequencies of the cavity
 were used in the calculations of the distribution $P(s)$.

Figure~2(a) shows that in spite of using short microwave antennas the experimental distribution $P(s)$ departures from the Poisson distribution (broken line)
 which is characteristic for classically integrable systems. The distribution $P(s)$ is also different from the theoretical prediction for Gaussian Orthogonal Ensemble (GOE) in RMT (full line)
characteristic for chaotic systems with time reversal symmetry, showing the transition between integrability and chaos. For $0.25\leq s \leq 0.75$ it is higher than the one for GOE in RMT, however, for  $s>2.0$ it is closer to the Poisson distribution.
This behavior is different than the one investigated by Robnik and Veble \cite{Robnik1998} for irrational and rational rectangles where huge fluctuations and the departure of the distribution $P(s)$  from the Poisson one were reported for very small $s$.

The results obtained for the microwave rectangular cavity should be contrasted with the ones obtained for the nearest neighbor spacing distribution $P(s)$ for the rough cavity (bars in Figure~2(b)) which shows much better agreement
with the theoretical prediction for GOE in RMT (full line).  In the case of the
rough cavity the distribution $P(s)$ was calculated on the basis of 3554 cavity eigenfrequencies. Some small discrepancies in the experimental $P(s)$ from the RMT prediction for $s \geq 2.25$ are possibly connected with either some unresolved resonances or fingerprints of nonuniversal behavior of the rough cavity. Similar discrepancies in the nearest neighbor spacing distribution $P(s)$ for $s \simeq 2.5$ are also visible in the experimental results presented in the paper by Poli et al. \cite{Poli2012}.
  Typical spectra of the rectangular cavity in the frequency range 16-17 GHz and the rough cavity in the frequency range 8-9 GHz are shown in Figure 3(a) and Figure 3(b), respectively.

\section{Experimental and numerical results for the elastic enhancement factor}

In Figure~4(a)  the elastic enhancement factor  $W$ of the two-port
scattering matrix $\hat{S}$
 of the microwave rectangular cavity is shown
  in a function of microwave frequency $\nu=16-18.5$ GHz (full circles). Due to significant fluctuations of the enhancement
  factor $W$ the experimental points were obtained by averaging of $W$ over 250 different realizations of the cavity
  length and the antennas positions in the frequency window $(\nu -\delta \nu/2, \nu +\delta \nu/2 )$, where $\delta \nu= 0.5$ GHz.
  The two black broken lines $W=2$ and $W=3$ show, respectively, the RMT limits for very strong and very low absorption.

The parameter $\gamma$ for the microwave rectangular cavities
depends on microwave frequency and was changed from 5.2 to 7.4
with the increase of frequency $\nu$ from 16 to 18.5 GHz.

According to Kharkov and Sokolov \cite{Kharkov2013}  the elastic enhancement factor  $W_{\beta}$ of the two-port
scattering matrix $\hat{S} $ evaluated within the framework of RMT can be expressed by

\begin{equation}
\label{Eq.3} W_{\beta}=2+\delta_{1 \beta}-\gamma \int_0^{\infty}d\tau e^{-\gamma \tau}b_{2,\beta}(\tau,\kappa ),
\end{equation}

where $b_{2,\beta}(\tau,\kappa )$ is the spectral form factor.
The parameter of chaoticity $\kappa$ changes from $\kappa=0$ for classically integrable systems to $\kappa \rightarrow\infty $ for chaotic systems.
It is important to note that in the transition region $0 < \kappa < \infty$ the spectral form factor $b_{2,\beta}(\tau,\kappa )$ is  currently known only for systems with broken time-reversal symmetry.
For integrable systems with time reversal symmetry $b_{2,1}(\tau,\kappa=0 )=0$ which immediately leads to $W=3$.

 Figure 4(a) shows that the elastic enhancement factor  $W$ of the two-port scattering matrix $\hat{S}$ of the rectangular cavity is below the theoretical value $W=3$. This result together with the complementary one for the experimental distribution $P(s)$ (see Figure 2(a))  strongly suggests that
 the system simulated by the two-port microwave rectangular cavity due to scattering on the antennas departures from the integrable one.
 This phenomenon was predicted by Seba \cite{Seba1990} and then thoroughly analyzed by Tudorovskiy et al. \cite{Tudorovskiy2010}.  The influence of antennas on the widths of resonances in a two-dimensional rectangular  microwave cavity, in much lower than investigated in this paper frequency range from below of 1 GHz to 5.5 GHz,  was studied by Barth\'elemy et al. \cite{Barthelemy2005}. In this experiment relatively short, 2 mm long, antennas were used. To give an idea about the antennas performance,  3 mm long antennas used is our experiment were characterized in the frequency range 16-18.5 GHz by the antenna coupling $\frac{1}{2}(T_a+T_b) \simeq 0.75$. In the frequency range 4.5-5.5 GHz, which was considered in \cite{Barthelemy2005}, the same antennas were characterized by much smaller antenna coupling $\frac{1}{2}(T_a+T_b) \simeq 0.15$.

 In order to estimate the chaoticity parameter $\kappa$ for such a system we reconstructed the nearest neighbor spacing distribution shown in Figure~2(a) using the random matrix Potter-Rosenzweig model described in  \cite{Leyvraz1991}, where the matrix $a_{ij}$ is defined as follows
\begin{equation}
\label{Eq.4}a_{ij}=g_{ij}[\delta_{ij} + \lambda (1-\delta_{ij})],
\end{equation}
where $g_{ij}$ denotes a symmetric matrix which belongs to GOE matrices. $\lambda$ is the transition
parameter. The off-diagonal elements $g_{ij}$ are independently Gaussian distributed with the same variance $\mbox{var}(g_{ij})=1$ and the mean zero. The diagonal elements  $g_{ii}$ are independently distributed  with the variance $\mbox{var}(g_{ii})=2$.

For the matrices $a_{ij}$ of the size $N\times N$ we found out that the parameter $\lambda$ can be approximated by $\lambda = \kappa/N$. The fit of the numerical nearest neighbor spacing distribution $P(s)$, calculated on the basis of 100 realizations of $200\times 200$ matrices, to the experimental one yields the chaoticity parameter $\kappa=2.8\pm 0.5$. The inset in Figure 2(a) shows the numerically reconstructed nearest neighbor spacing distribution $P(s)$.
Unfortunately, even knowing the chaoticity parameter $\kappa$ we are not able to compare our experimental results with the theoretical ones since the explicit form of the spectral form factor $b_{2,1}(\tau,\kappa )$ is not known. Though the paper \cite{Kharkov2013} suggests that the behaviour of the enhancement factor for systems with time reversal symmetry should be similar to the one for systems with broken time reversal symmetry, this remains to be proven yet. Just for completeness of the presentation in Figure~4(a) we also show the RMT results predicted by the equation (3) (empty cirles) with the spectral form factor $b_{2,1}(\tau,\kappa \rightarrow \infty )$ defined by the equation (6).

  In Figure~4(b)  the elastic enhancement factor  $W$ of the two-port
scattering matrix $\hat{S}$
 of the microwave rough cavity simulating a quantum chaotic system is shown (full black rhombi)
  in a function of microwave frequency $\nu=16-18.5$ GHz. The results were averaged over 105 different perturber positions
  in the frequency window $\delta \nu= 0.5$ GHz.
  It is important to note that the theoretical and experimental investigations of rough billiards (cavities) \cite{Fram1997,Hlushchuk2001,Savytskyy2004} showed  that for
lower energies (frequencies) there exist regimes of localization and Wigner ergodicity, and only for higher
energies (frequencies) billiards (cavities) become fully chaotic. This fully chaotic regime is called the regime of Shnirelman ergodicity.
 For the rough cavity used in the experiment the regime of Shnirelman ergodicity extends for $\nu > 9.9$ GHz. The presence of the perturber causes that even for lower frequencies $\nu=6-9$ GHz the nearest neighbor spacing distribution $P(s)$ is close to the theoretical prediction for GOE in RMT.

   The parameter  $\gamma= \frac{1}{2}(\gamma^{(a)}+\gamma^{(b)})$  was
estimated by adjusting the
theoretical mean reflection coefficients parameterized  by the parameters $\gamma^{(j)}$

\begin{equation}
\label{Eq.5} \langle R \rangle _{th}^{(j)} = \int _0^1dRRP(R),
\end{equation}
 to the experimental ones $\langle R \rangle^{(j)}$ obtained after eliminating the direct
processes  \cite{Anlage2006,Lawniczak2008}. The index $j=a,b$ denotes the
port $a$ or $b$.
In the calculations of $\langle R \rangle _{th}^{(j)} $ we used the analytic expression for the distribution $P(R)$ of the
reflection coefficient $R$  given in \cite{Savin2005}.
We found out that using the same antennas, 3 mm long, as in the case of the rectangular cavity, the value of the parameter $\gamma$ was changed from 5.3 to 6.8 with the increase
of  microwave frequency $\nu$ from 16 GHz to 18.5 GHz, respectively. Taking into account that the microwave antennas
act as single scattering channels the absorption strength $\gamma$ can be expressed as a sum of the transmission coefficients: $\gamma =\sum_c T_c=T_a+T_b+\alpha $, where $\alpha$  represents internal absorption of the cavity \cite{Dietz2010}.  The values of the absorption strength $\gamma $ and the transmissions coefficients $T_a$, $T_b$, and $\alpha$ are shown in Table \ref{tab:abs}.

\begin{table}[!hbt]
\centering
\begin{tabular}{|l|l|l|l|l|}
 \hline
 $\delta \nu \mbox{/GHz }$ & $\gamma$ & $T_a$ & $T_b$ & $\alpha$ \\
 \hline
 $16.0-16.5$ & $5.32$ & $0.57$ & $0.61$ & $4.14$\\
 \hline
 $16.5-17.0$ & $5.82$ & $0.67$ & $0.64$ & $4.51$\\
 \hline
 $17.0-17.5$ & $6.29$ & $0.61$ & $0.78$ & $4.90$\\
 \hline
 $17.5-18.0$ & $6.55$ & $0.73$ & $0.84$ & $4.98$\\
 \hline
 $18.0-18.5$ & $6.82$ & $0.74$ & $0.84$ & $5.24$\\
 \hline
\end{tabular}
\caption{The absorption strength $\gamma$, the transmissions coefficients $T_a$, $T_b$  and the internal
absorption of the cavity $\alpha$ in the frequency range $\delta \nu$.}
\label{tab:abs}
\end{table}

Figure~4(b) shows that the experimental results obtained for the rough cavity are below the theoretical ones predicted for $W$  by the equation (3) within the framework of RMT (empty rhombi). For chaotic systems ($\kappa \rightarrow \infty$) with the symmetry index $\beta=1$ the spectral form factor $b_{2,1}(\tau,\kappa \rightarrow \infty )$ in the equation (3) has the form \cite{Fyodorov2005,Savin2006}:

\begin{equation}
\label{Eq.6}
b_{2,1}(\tau, \kappa \rightarrow \infty ) = [1 -2\tau + \tau \log(1 + 2\tau )]\Theta(1 - \tau )
+[\tau \log\frac{2\tau +1}{2\tau-1}-1]\Theta(\tau-1),
\end{equation}

where $\Theta(\cdot)$ is the Heaviside step function.

It is important to point out that the recent numerical results presented in \cite{Sokolov2014} for the two-channel problem with absorption give better agreement with the experimental ones. In the case of the two equivalent channels with $0.6\leq T\leq 0.8$ and the internal absorption $\alpha =3$ the theory predicts $W$ to be between 2.08 and 2.03 (see Figure 5 in \cite{Sokolov2014}). In the experiment the internal absorption $\alpha \simeq 5$ was larger than the ones considered in the theoretical calculations therefore one should expect even smaller theoretical values of the elastic enhancement factor $W$.
For comparison, the experimental elastic enhancement factor $W$  is scattered between 2.1 and 1.95.

\section{Conclusions}

The elastic enhancement factor
$W$ was experimentally studied  for microwave rectangular and rough cavities simulating
 partially chaotic, characterised by the transient parameter $\kappa = 2.8$, and chaotic two-dimensional quantum billiards, respectively. Both systems were characterized by similar, moderate absorption strengths, $\gamma=5.2-7.4$ and $\gamma=5.3-6.8$, respectively.
 We show that the results obtained for the rectangular cavity lie below the theoretical prediction for integrable systems  $W=3$, however, they are significantly higher than the ones obtained for the microwave rough cavity. The results obtained for the microwave rough cavity are smaller than the ones obtained within the framework of RMT and lie between them and the ones predicted within a model of the two-channel coupling recently introduced by Sokolov and Zhirov \cite{Sokolov2014}.
   Our experimental results suggest that the elastic enhancement factor can be used as a measure of internal chaos that can be especially useful for systems with significant absorption or openness.

\section{Acknowledgments}

 We are very grateful to V. Sokolov for fruitful discussions. This work was partially supported by the Ministry of Science and Higher Education grants N N202 130239 and UMO-2013/09/D/ST2/03727.

\pagebreak


\smallskip

\begin{figure}[tb]
\begin{center}
\rotatebox{0}{\includegraphics[width=0.8\textwidth,
height=0.6\textheight, keepaspectratio]{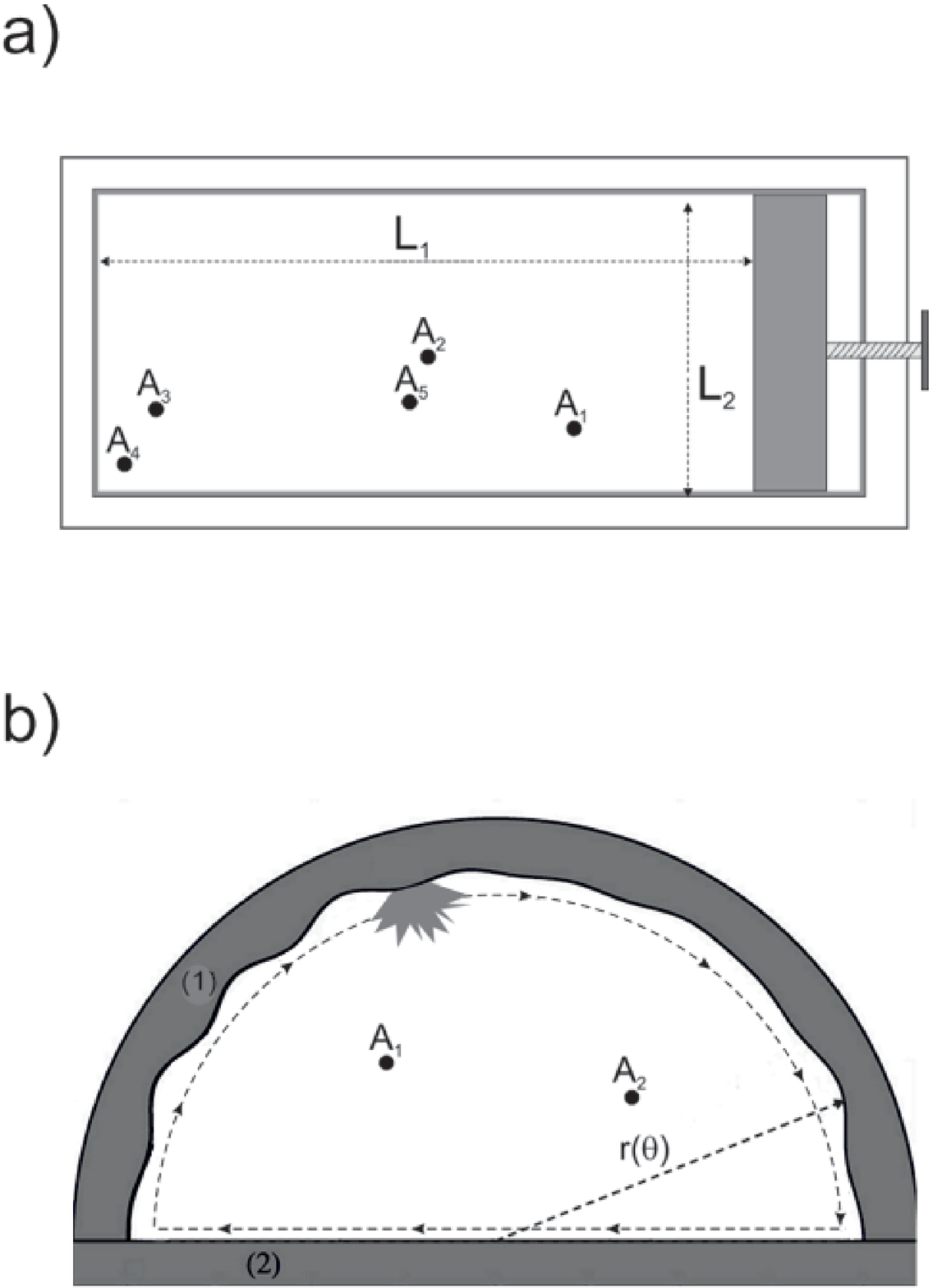}}
 \caption{(a) The rectangular microwave cavity and (b) the rough cavity which were
used for measuring of the two-port scattering matrix $\hat{S}$.
The rough cavity side wall segments are marked by (1) and (2) (see text).
The scattering matrix $\hat{S}$ of the cavities were measured in
the frequency window: 16--18.5 GHz. The vector network analyzer
Agilent E8364B was connected to the microwave antennas which were
introduced inside the cavities (holes $A_1$, $A_2$, $A_3$, $A_4$, $A_5$
in Figure~1(a) and $A_1$, $A_2$ in Figure~1(b) ) through the
flexible microwave cables HP 85133-616 and HP 85133-617. The width
of the rectangular cavity was 20 cm. The length of the cavity was
changed from $L_1 = 41.5$ to 36.5 cm in 25 steps of 0.2 cm length.
In order to create different realizations of the rough cavity a
metallic perturber (see panel (b)) was moved inside the cavity.
}\label{Fig1}
\end{center}
\end{figure}

\begin{figure}[tb]
\begin{center}
\rotatebox{0}{\includegraphics[width=0.8\textwidth,
height=0.6\textheight, keepaspectratio]{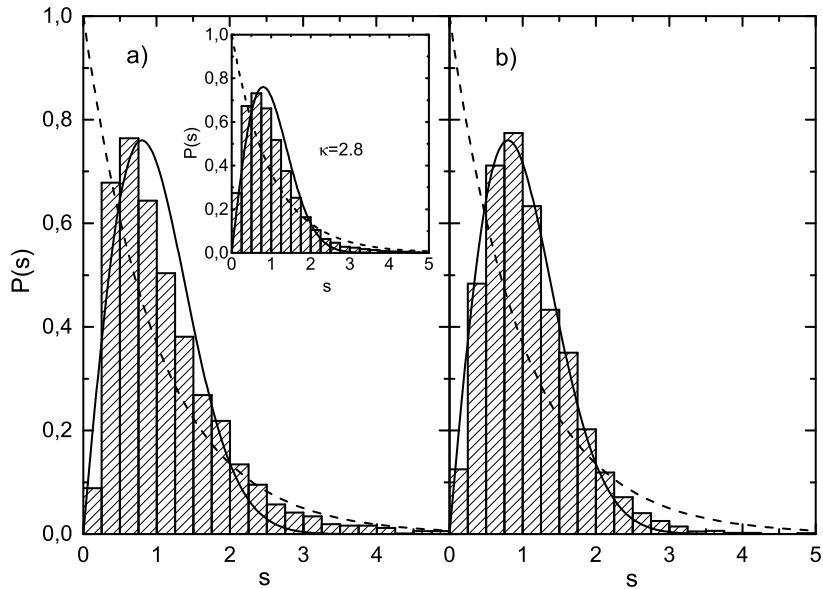}} \caption{(a) The
nearest neighbor spacing distribution $P(s)$ for the microwave rectangular
cavity coupled to the external channels via antennas (bars) obtained for the frequency range $\nu=16-17$ GHz.
The experimental distribution $P(s)$
is compared to the Poisson distribution (broken line) which is
characteristic for classically integrable systems and to the theoretical
prediction for GOE in RMT (full
line). The inset shows
the numerically reconstructed nearest neighbor spacing distribution $P(s)$ for the chaoticity parameter $\kappa=2.8$.
(b) The nearest neighbor spacing distribution $P(s)$ for
the microwave rough cavity obtained for the frequency range $\nu=6-9$ GHz
 (bars) is compared to the theoretical prediction for GOE (full line) and to the Poisson distribution (broken line).
}\label{Fig2}
\end{center}
\end{figure}

\begin{figure}[tb]
\begin{center}
\rotatebox{0}{\includegraphics[width=0.8\textwidth,
height=0.6\textheight, keepaspectratio]{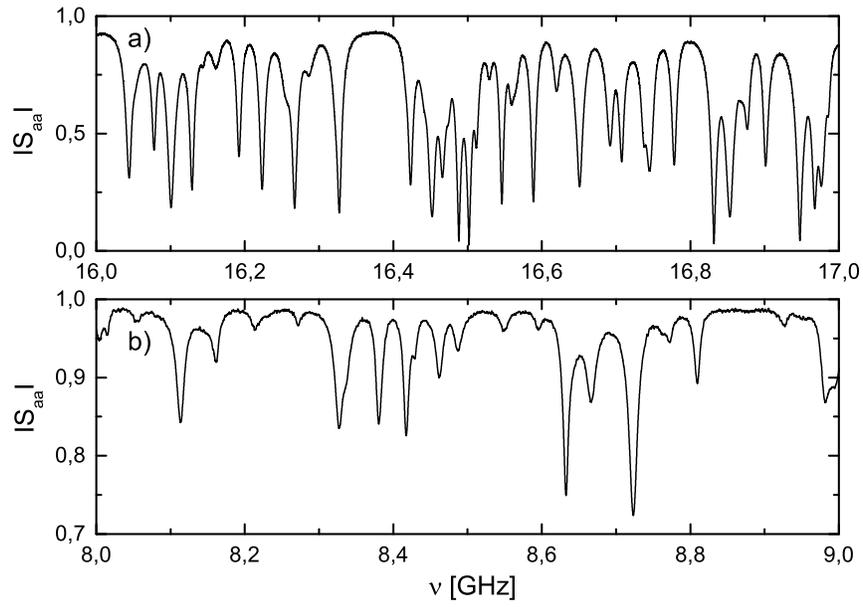}} \caption{Typical spectra of the rectangular cavity in the frequency range 16-17 GHz (panel (a)) and the rough cavity in the frequency range 8-9 GHz (panel (b)).}\label{Fig3}
\end{center}
\end{figure}

\begin{figure}[tb]
\begin{center}
\rotatebox{0}{\includegraphics[width=0.9\textwidth,
height=0.7\textheight, keepaspectratio]{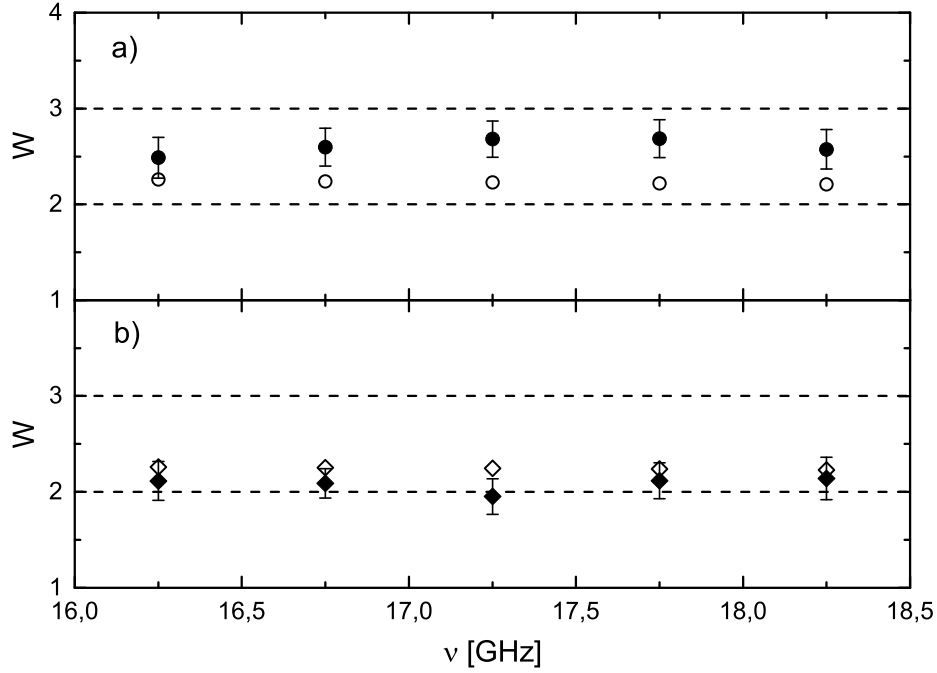}}
 \caption{(a)
The elastic enhancement factor  $W$ of the two-port scattering matrix
$\hat{S}$ of the rectangular cavity coupled to the external channels via antennas simulating
a quantum system with the chaoticity parameter $\kappa=2.8\pm 0.5$ (full circles).
 The RMT results \cite{Fyodorov2005,Savin2006} are shown by empty circles.
  (b) The elastic enhancement factor  $W$ of the two-port
scattering matrix $\hat{S}$
 of the microwave rough cavity simulating a quantum chaotic system (full black rhombi).
 The RMT results are shown by empty rhombi.
The measurements were done in the frequency window $\nu=16-18.5$ GHz.
The two black broken lines $W=2$ and $W=3$ show, respectively,  the RMT limits
for very strong and very low absorption. The latter limit $W=3$ is also expected for the integrable systems.
}\label{Fig4}
\end{center}
\end{figure}

\end{document}